\documentclass{PoS}

\usepackage[T1]{fontenc}
\usepackage[utf8]{inputenc}
\usepackage{amsmath}
\usepackage{amssymb}
\usepackage{tikz}
\usepackage[outercaption]{sidecap}

\IfFileExists{../../../../Dokumente/Mendeley_Desktop/library.bib}{%
	\newcommand{\Bibliography}{../../../../Dokumente/Mendeley_Desktop}}{%
	\newcommand{\Bibliography}{/home/t30/wei/ga56gix/Dokumente/Mendeley_Desktop}
	}

\newcommand{\als}{\alpha_{\text{s}}}
\newcommand{\Nt}{N_{\tau}}
\newcommand{\mD}{m_{\text{D}}}
\newcommand{\Nc}{N_{\text{c}}}
\newcommand{\Nf}{N_{\text{f}}}
\newcommand{\CF}{C_{\text{F}}}
\newcommand{\gammaE}{\gamma_{\text{E}}}

\newcommand{\tr}{\text{tr}}

\definecolor{Mblue}{rgb}{0.368417, 0.506779, 0.709798}
\definecolor{Morange}{rgb}{0.880722, 0.611041, 0.142051}
\definecolor{Mgreen}{rgb}{0.560181, 0.691569, 0.194885}
\definecolor{Mred}{rgb}{0.922526, 0.385626, 0.209179}
\definecolor{Mviolet}{rgb}{0.529, 0.471, 0.702}
\definecolor{Mgray}{rgb}{0.3, 0.3, 0.3}

\title{Color screening in $2+1$ flavor QCD at large distances}

\ShortTitle{Color screening in $2+1$ flavor QCD at large distances}

\author{\speaker{Sebastian Steinbeißer}\\
	Physik Department\\
    Technical University of Munich\\
    James-Franck-Str. 1\\
    85748 Garching, Germany\\
    E-mail: \email{sebastian.steinbeisser@tum.de}}
\author{Johannes Heinrich Weber\\
	Department of Computational Mathematics, Science and Engineering (CMSE)\\
	Michigan State University\\
	E-mail: \email{weberjo8@msu.edu}}
\author{(for TUMQCD Collaboration)}

\abstract{
We study correlation functions of spatially separated static quark-antiquark 
pairs in $2+1$ flavor QCD in order to investigate the nature of color screening 
at high temperatures. We perform lattice calculations in a wide temperature 
range, $116~\text{MeV} \leq T \leq 5814~\text{MeV}$, using the highly improved 
staggered quark (HISQ) action and several lattice spacings to control 
discretization effects. We alleviate the UV noise problem through the use of 
four dimensional hypercubic (HYP) smearing, which enables the reconstruction of 
correlators and determination of screening properties even at low temperatures 
and at large distances.\\
This proceeding contribution is a report on work in progress aimed at improving 
previous results~\cite{Bazavov:2018wmo} and prepares a forthcoming 
publication~\cite{paper2}.
}

\FullConference{XIII Quark Confinement and the Hadron Spectrum - 
	Confinement2018\\
	31 July - 6 August 2018\\
	Maynooth University, Ireland}

\begin{document}

\section{Introduction}

At zero temperature a heavy quark and antiquark pair forms a bound state 
(quarkonium state) or a pair of heavy-light mesons. At sufficiently high 
temperatures ($T > T_{c}$), (heavy) quarks are not confined anymore due to 
color screening and collisions with the medium. The Polyakov loop and the 
Polyakov loop correlator are the order parameters of this 
confinement-deconfinement transition from hadrons to a quark gluon plasma (QGP) 
in pure Yang-Mills theory and still provide sensitive probes in full QCD. The 
continuum limit of the corresponding quark-antiquark free energies have been 
studied previously (see Ref.~\cite{Bazavov:2018wmo}) and it has been found that 
their short distance behavior can be understood in terms of weak-coupling 
effective field theory calculations in the framework of pNRQCD and EQCD (see 
Ref.~\cite{Berwein:2017thy}). These studies have been limited by the 
exponential drop of the signal to noise ratio, which we can now alleviate 
through the use of smeared gauge fields. We demonstrate that through 
subtraction of the squared expectation value of the appropriately smeared 
Polyakov loop we can reconstruct the subtracted free energies at sufficiently 
large distances and smoothly connect results with different amounts of 
HYP-smearing (see Ref.~\cite{Hasenfratz:2001hp}) applied. We use four 
dimensional HYP-smearing throughout this work. This allows a quantitatively 
predictive study even in the asymptotic screening regime.\\
We perform lattice calculations in a wide temperature range, $116~\text{MeV} 
\leq T \leq 5814~\text{MeV}$, using the highly improved staggered quark (HISQ) 
action (see Ref.~\cite{Follana:2006rc}) and the MILC code (see 
Ref.~\cite{Bazavov:2010ru}) and several lattice spacings in 2+1 flavor QCD.\\
The $Q\bar{Q}$ color average and color singlet (subtracted) free energies are 
given in terms of the respective (subtracted) Polyakov loop correlators by
\begin{equation}
\label{eq:FreeEnergies}
F_{Q\bar{Q}}^{(\text{sub.})} = -T \ln(C_{P}^{(\text{sub.})}) \,, \quad\quad 
F_{\text{s}}^{(\text{sub.})} = -T \ln(C_{\text{s}}^{(\text{sub.})}) \,.
\end{equation}
The bare (unsubtracted) Polyakov loop correlators are given by
\begin{equation}
C_{P}(r) = \langle P(0) P^{\dagger}(r) \rangle \,, \quad\quad C_{\text{s}}(r) = 
\frac{1}{3} \langle \tr W(0) W^{\dagger}(r) \rangle \,.
\end{equation}
The Polyakov loop $P$ being the trace of a temporal Wilson line $W$, where the 
latter is not gauge invariant; we define $C_{\text{s}}$ in Coulomb gauge.
\begin{equation}
P(r) = \frac{1}{3} \tr W(r) \,, \quad\quad W(r) = \prod\limits_{\tau/a = 
1}^{N_{\tau}} U_{0}(\tau,r) \,,
\end{equation}
with $W$ describing the propagation of a static quark via the link variables 
$U_{0}$. The expectation value of the Polyakov loop averaged over the spatial 
volume of the lattice, $L = \langle P(r) \rangle$, is used to normalize 
Polyakov loop correlators
\begin{equation}
\label{eq:Normalization}
C_{P}^{\text{sub.}}(r) = \frac{C_{P}(r)}{L^{2}} \,, \quad\quad 
C_{\text{s}}^{\text{sub.}}(r) = \frac{C_{\text{s}}(r)}{L^{2}} \,.
\end{equation}
This normalizes the correlator $C^{\text{sub.}}$ such that the corresponding 
free energy $F^{\text{sub.}}$ vanishes at infinite separation in infinite 
volume. It is important to note that the correlator and $L$ need to be computed 
for the same regularization, i.e., with the same HYP-smearing level. The 
subtracted free energies defined in terms of the subtracted correlators are 
divergence free and have a well-defined continuum limit. At distances $r 
\gtrsim 1/(gT)$, where $gT \sim \mD$ is the Debye mass, the dimensionally 
reduced EFT called electrostatic QCD (EQCD) is suitable to describe these 
correlators (see Ref.~\cite{Braaten:1995jr}). For distances $r \ll 1/(gT)$, the 
correlators are sensitive to the inherent non-perturbative magnetic sector and 
receive contribution from the magnetic mass $\propto g^{2}T$. Hence, a 
non-perturbative approach is required for studying the large distance 
behavior.

\section{Screening functions for color singlet and color average free energies}

\begin{figure}[h]
\centering
\includegraphics[width=0.49\textwidth]{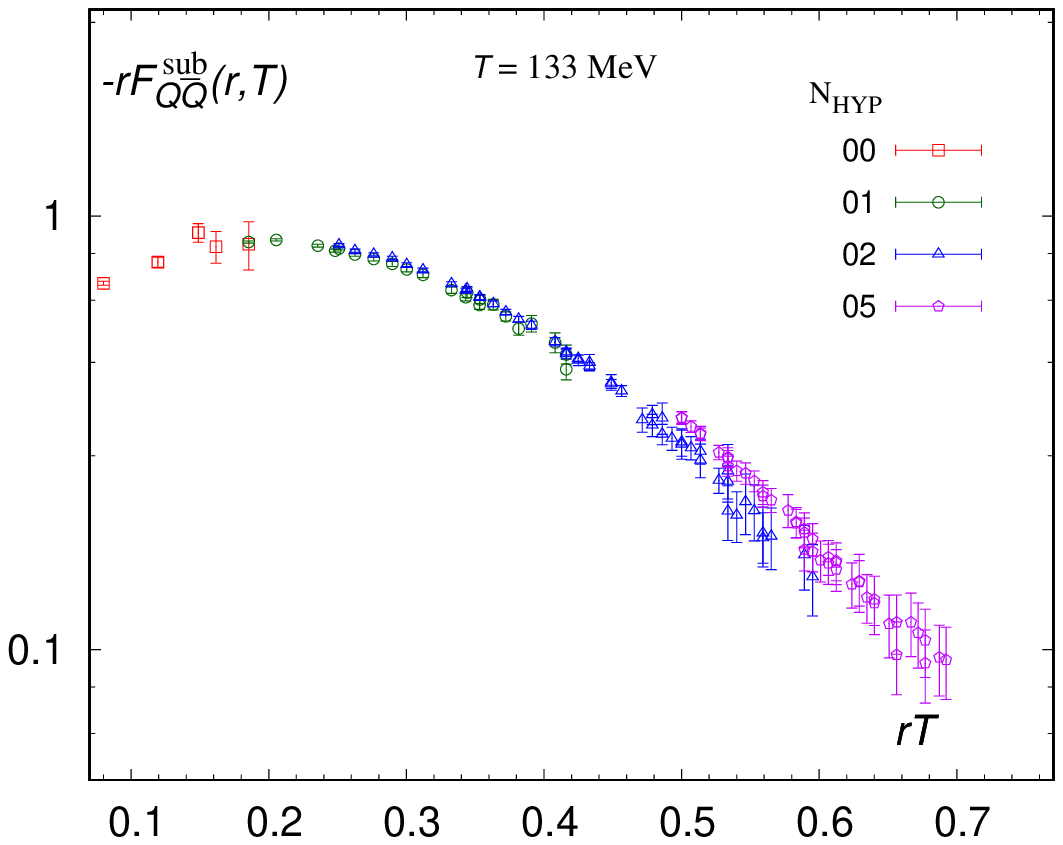}
\hfill
\includegraphics[width=0.49\textwidth]{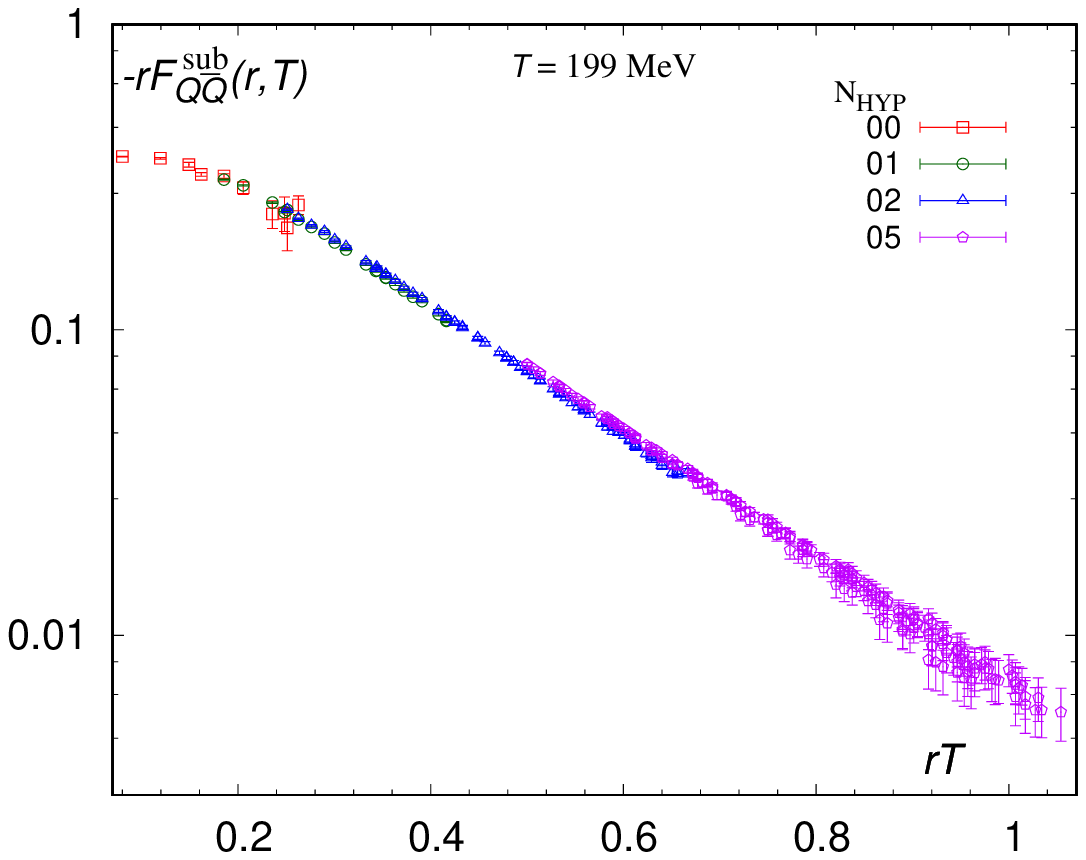}
\caption{Screening functions for color average free energies. We show results 
for $\Nt=12$ below $T_{c}$ (left panel) and above $T_{c}$ (right panel) for 
different HYP-smearing levels. These blend nicely into each other and, as 
discussed below, this allows us to produce a screening function spanning a 
large $rT$ range.}
\label{fig:ScreeningFunctionsI}
\end{figure}
In the previous study~\cite{Bazavov:2018wmo} it was shown that data with 
$\Nt=12$ is only marginally different from the continuum limit already. We use 
these as a proxy for a full continuum limit in the following. Multiplying the 
subtracted free energies for the color singlet and color average in units of 
the temperature $T$ by $-r$, we obtain the respective dimensionless screening 
functions assuming one-particle exchange. In Fig.~\ref{fig:ScreeningFunctionsI} 
we show results for different HYP-smearing levels for two temperatures below 
and above $T_{c}$, respectively, for $\Nt=12$. It is remarkable how well data 
at different smearing levels blend into one another allowing us to reach $rT$ 
regions well above $rT \sim 0.4$ where color screening is supposed to set in. 
For $T < T_{c}$ the noise exceeds the unsmeared signal already before the 
screening is unambiguously visible. Smearing allows us to piece together a 
combined screening function covering a substantially larger $rT$ range than in 
the earlier analysis~\cite{Bazavov:2018wmo}, especially for $T \lesssim 2T_{c}$.
\begin{figure}[h]
\centering
\includegraphics[width=0.49\textwidth]{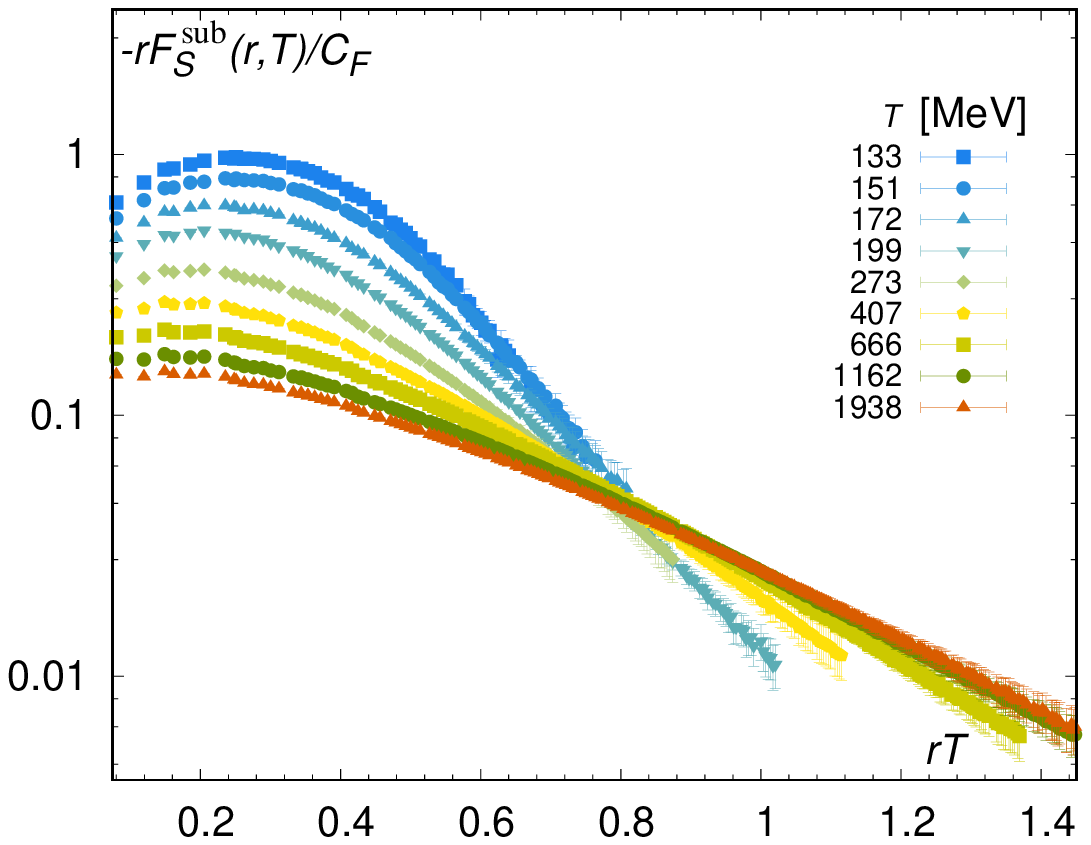}
\hfill
\includegraphics[width=0.49\textwidth]{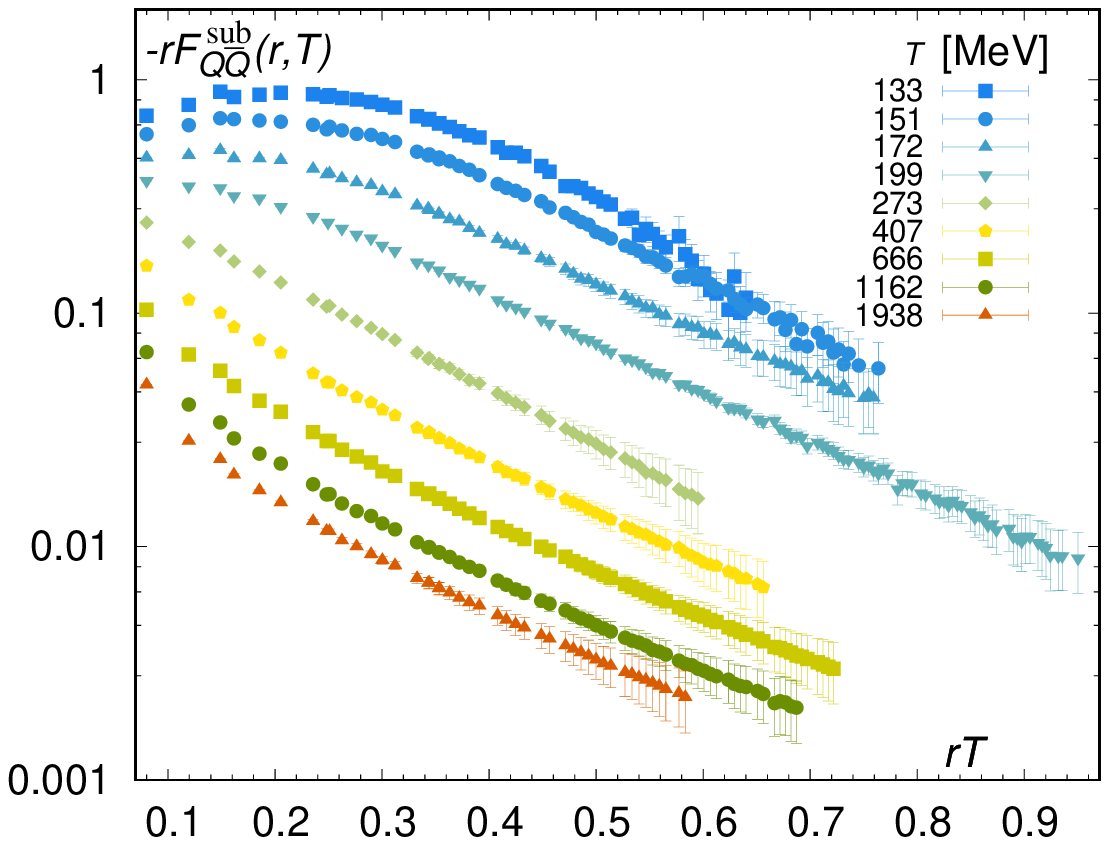}
\caption{Screening functions for the color singlet (left panels) and the color 
average free energies (right panels). We show results for $\Nt=12$ for a 
large temperature range.}
\label{fig:ScreeningFunctionsII}
\end{figure}
These aggregate screening functions are shown in 
Fig.~\ref{fig:ScreeningFunctionsII}, where we show 
$-rF_{\text{s}}^{\text{sub.}}$ and $-rF_{Q\bar{Q}}^{\text{sub.}}$ with $\Nt=12$ 
over a large temperature range. Both should appear as straight lines in the 
log-plot for naive exponential screening $\propto \exp(-mr)$. This is indeed 
the case, indicating that at sufficiently large distances both 
$F_{Q\bar{Q}}^{\text{sub.}}$ and $F_{\text{s}}^{\text{sub.}}$ decay 
exponentially. In contrast to the former analysis (see 
Ref.~\cite{Bazavov:2018wmo}), we can see this now even for small $T$, where the 
noise outgrows the signal without smearing before the exponential decay even 
sets in. HYP-smearing allows us to extend the signal up to $rT \sim 0.8$ in 
order to see and also fit the exponential decay.\\
As a general feature we see that the color singlet screening function gets 
steeper as $rT$ increases but flatter with increasing $T$. On the other hand 
the color average screening function changes its short distance behavior before 
entering the screening regime. This is due to the change from a Coulombic 
behavior in the regime $T \ll \als/r$ to a $1/r^{2}$ behavior in the regime $T 
\gg \als/r$, which can be understood as due to cancellation between color 
singlet and color octet contributions as seen in weak coupling calculations. 
For the screening regime, we find that the color average screening function 
does not exhibit distinct features, i.e., the slope does not change 
significantly with increasing $T$ or $rT$. The leading order result for the 
color average free energy in the color electric screening regime is given by 
the exchange of two electro-static gluons (see 
Refs.~\cite{Gross:1980br,McLerran:1981pb,Nadkarni:1986cz}):
\begin{equation}
F_{Q\bar{Q}}^{\text{sub.}} \simeq -\frac{\als^{2}}{r^{2}T} \exp(-2\mD r) \,.
\end{equation}
The perturbative NLO Debye mass (see Ref.~\cite{Braaten:1995jr}) in temperature 
units is given by:
\begin{align}
\begin{aligned}
\label{eq:DebyeMass}
& \mD|_{\text{LO}}(\mu) = g(\mu) T \sqrt{\frac{2\Nc + \Nf}{6}} \,, \\
& \mD^{2}|_{\text{NLO}}(\mu) = \mD^{2}|_{\text{LO}}(\mu) \Big( 1 + 
\frac{\als(\mu)}{4\pi} \Big[ 2 \beta_{0} \left( \gammaE + \ln \frac{\mu}{4 \pi 
T} \right) \\
& \quad\quad\quad\quad\quad + \frac{5\Nc}{3} + \frac{2\Nf}{3} (1 - 4 \ln 2) 
\Big] \Big) - \CF \Nf \als^{2}(\mu) T^{2} \,.
\end{aligned}
\end{align}
\begin{figure}[h]
\centering
\includegraphics[width=0.49\textwidth]{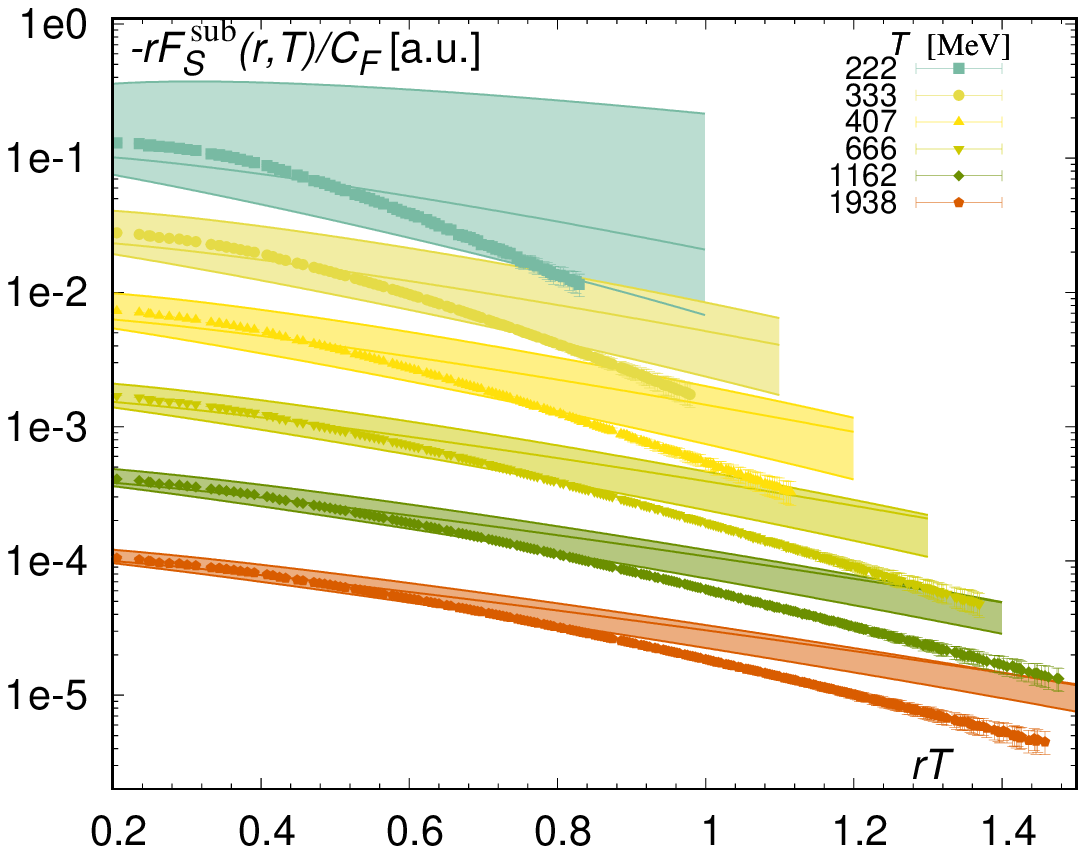}
\hfill
\includegraphics[width=0.49\textwidth]{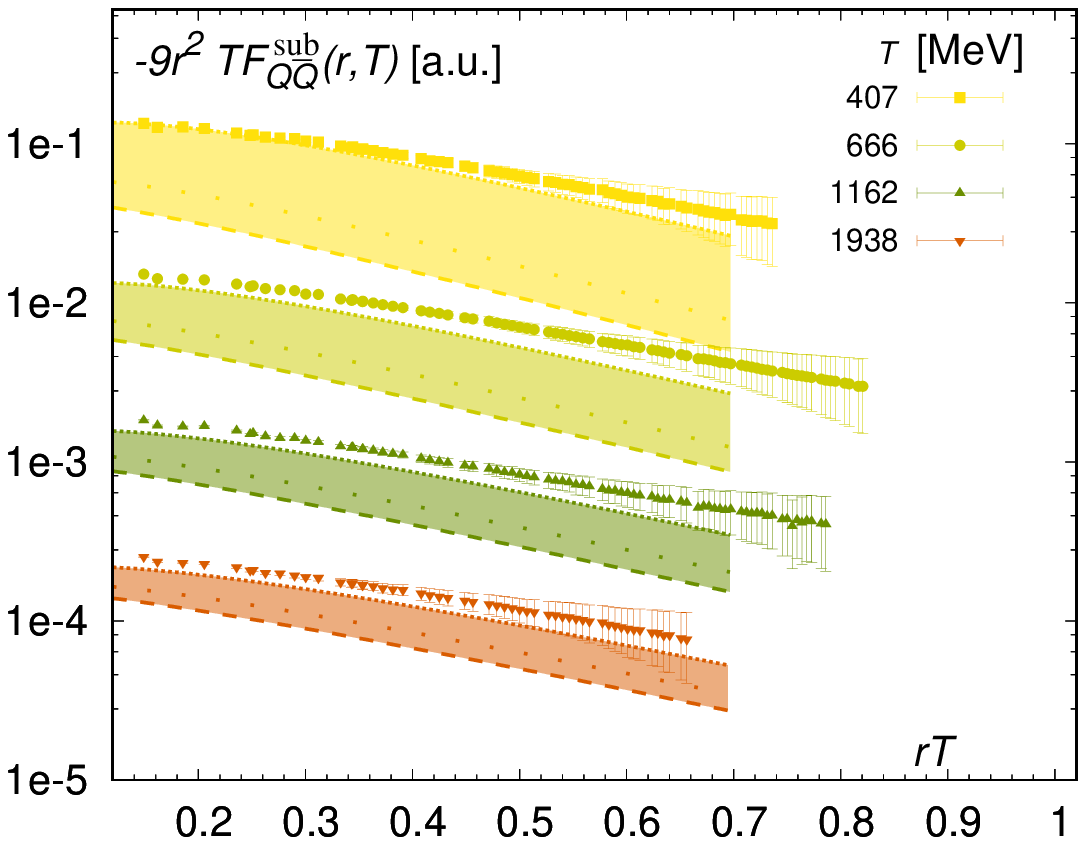}
\caption{Comparison of the screening functions for the color singlet (left 
panel) and the color average free energies (right panel) to EQCD predictions 
(bands) at NLO. The widths of the bands correspond to the variation of the 
scale $\mu$ between $\pi T$ and $4\pi T$. We show results for $\Nt=12$. The 
results for different $T$ have been vertically displaced for better visibility.}
\label{fig:ScreeningFunctionsIV}
\end{figure}
In the same regime of color electric screening, $r \sim 1/\mD$, also the color 
singlet free energy can be described by EQCD (See Eq.~(21) in 
Ref.~\cite{Bazavov:2018wmo} for full formulae). A comparison between both 
screening functions and EQCD predictions using the NLO Debye mass given in 
Eq.~\eqref{eq:DebyeMass} is shown in Fig.~\ref{fig:ScreeningFunctionsIV}. The 
band reflects the variation of the scale $\mu$ between $\pi T$ and $4\pi T$.\\
At larger distances $r \gg 1/\mD$, contributions from the magnetic scale 
$\propto g^{2}T$ may be relevant and, thus, a non-perturbative calculation is 
required. We expect an asymptotic screening behavior as $F^{\text{sub.}} \sim 
\exp(-mr)/r$. For $F_{Q\bar{Q}}$, this is due to the exchange of a single bound 
state of electric gluons, which may mix with contributions from the magnetic 
sector. We expect a similar behavior for the color singlet free energy. 
Comparing the $\Nt=12$ results for $F_{Q\bar{Q}}$ that are shown in the 
Figs.~\ref{fig:ScreeningFunctionsII} and \ref{fig:ScreeningFunctionsIV}, we see 
that mainly due to the insensitivity of the screening function to changes in 
$rT$, a distinction between the color electric screening regime and the 
asymptotic screening regime is difficult. This should manifest itself in a 
nearly $rT$ independent screening mass. The EQCD prediction contains higher 
order corrections due to the running coupling that are numerically important 
and can be seen in the data. The picture for the color singlet is different, as 
we see the increase in slope for increasing $rT$. This signals the onset of the 
asymptotic screening regime. We expect a significant $rT$ dependence of the 
screening mass.

\section{Temperature dependence of the color singlet and color average screening mass}

\begin{SCfigure}[\sidecaptionrelwidth][h]
\centering
\includegraphics[width=0.49\textwidth]{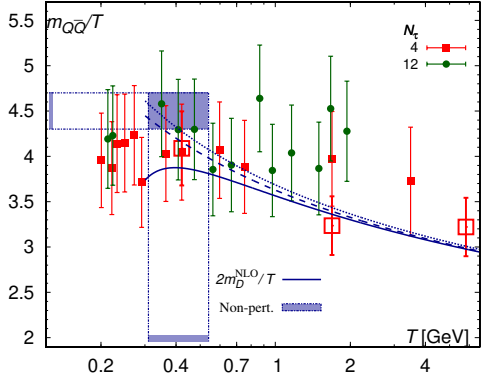}
\caption{Color average asymptotic screening masses obtained from the 
exponential fits to the subtracted screening functions. The lines correspond to 
two times the NLO Debye mass in temperature units which is calculated for 
$\mu=\pi T$, $2\pi T$, and $4\pi T$ (solid, dotted and dashed lines). The red 
open squares correspond to the screening masses using $\Nt=4$ lattices with 
aspect ratio 6. The horizontal band in the right panel corresponds to the EQCD 
result for the screening mass obtained in 3d lattice calculations from 
Ref.~\cite{Hart:2000ha}.}
\label{fig:ScreeningMass}
\end{SCfigure}
We perform several different fits of the screening functions in order to obtain 
screening masses. The generic fit function for both the color singlet and color 
average screening function is of the form
\begin{equation}
\label{eq:FitRF}
-RF = (A \exp(-MR) + cR) \Theta(R - i/N_{\tau}) \,.
\end{equation}
where $M=m/T$, $R=rT$, and $F$ is given in units of $T$. The fit parameters are 
$A$, $M$, and $c$, where for the subtracted free energies $c$ is supposedly 
zero in the infinite volume limit, but may actually be non-negligible due to 
finite volume effects or improper cancellation for even moderate statistics. In 
order to take into account that HYP-smearing distorts UV physics, we demand the 
minimal value of $R$ to be at least the number of smearing iterations divided 
by $\Nt$. This is ensured by the Heaviside step function. For $\Nt=4$ with 
aspect ratio 4 we consider 0,1,2, and 3 HYP-smearing iterations, i.e. $i \in 
\lbrace 0,1,2,3 \rbrace$, and for $\Nt \in \lbrace 6,8,10,12,16 \rbrace$ and 
for $\Nt=4$ with aspect ratio 6 we additionally consider 5 HYP-smearing 
iterations, i.e. $i \in \lbrace 0,1,2,3,5 \rbrace$. We determine for each 
ensemble and HYP-smearing iteration a sensible maximal value for $R$, beyond 
which we no longer see exponential decay or the uncertainties of the data 
become too large. Within these bounds of $R$ we vary $R_{\text{min}}$.\\
In a first approach we treat all the different ensembles and all smearing 
iterations on each ensemble as independent and perform the fits on the 
Jackknife averages of our data. This approach, however, lacks a handle on 
systematic uncertainties. In order to overcome this we perform individual fits 
according to Eq.~\eqref{eq:FitRF} on the Jackknife bins of our data.\\
Figure~\ref{fig:ScreeningMass} shows the asymptotic color average screening 
mass $m_{Q\bar{Q}}/T$ in temperature units obtained at $rT=0.5$ from the 
exponential fits of the screening function performed on the Jackknife bins. The 
results have been shifted by $-0.25$ with a systematic error estimate of $\pm 
0.1$. The red open squares are at $rT=1.3$ and correspond to the screening 
masses using $\Nt=4$ lattices with aspect ratio 6. The lines correspond to 
twice the EQCD prediction, Eq.~\eqref{eq:DebyeMass}, for $\mu = \pi T$, $2\pi 
T$ and $4\pi T$. The similarity of the asymptotic mass $m_{Q\bar{Q}}$ to the 
electric screening mass in EQCD $2\mD$ makes a distinction between the two 
regimes for $F_{Q\bar{Q}}$ particularly difficult. Our result is in good 
agreement with a lattice determination of the EQCD screening mass (see 
Ref.~\cite{Hart:2000ha}), shown as a horizontal band.\\
For the color singlet (not shown) we are able to obtain results at $rT=1$. 
Using a $\Nt=4$ aspect ratio 6 determination at $rT=1.3$, we need to shift the 
$rT=1$ results by $+0.15$ with a systematic error estimate of $\pm 0.05$ in 
order to be consistent with the previous results. The corresponding EQCD 
prediction, Eq.~\eqref{eq:DebyeMass}, for $\mu = \pi T$, $2\pi T$ and $4\pi T$, 
requires a rescaling by a constant $A=1.6-2.0$. We then obtain that for $T 
\gtrsim 400~\text{MeV}$ the screening mass and the rescaled Debye mass are very 
similar.

\section{Temperature dependence of asymptotic screening masses}

\begin{figure}[h]
\centering
\includegraphics[width=0.49\textwidth]{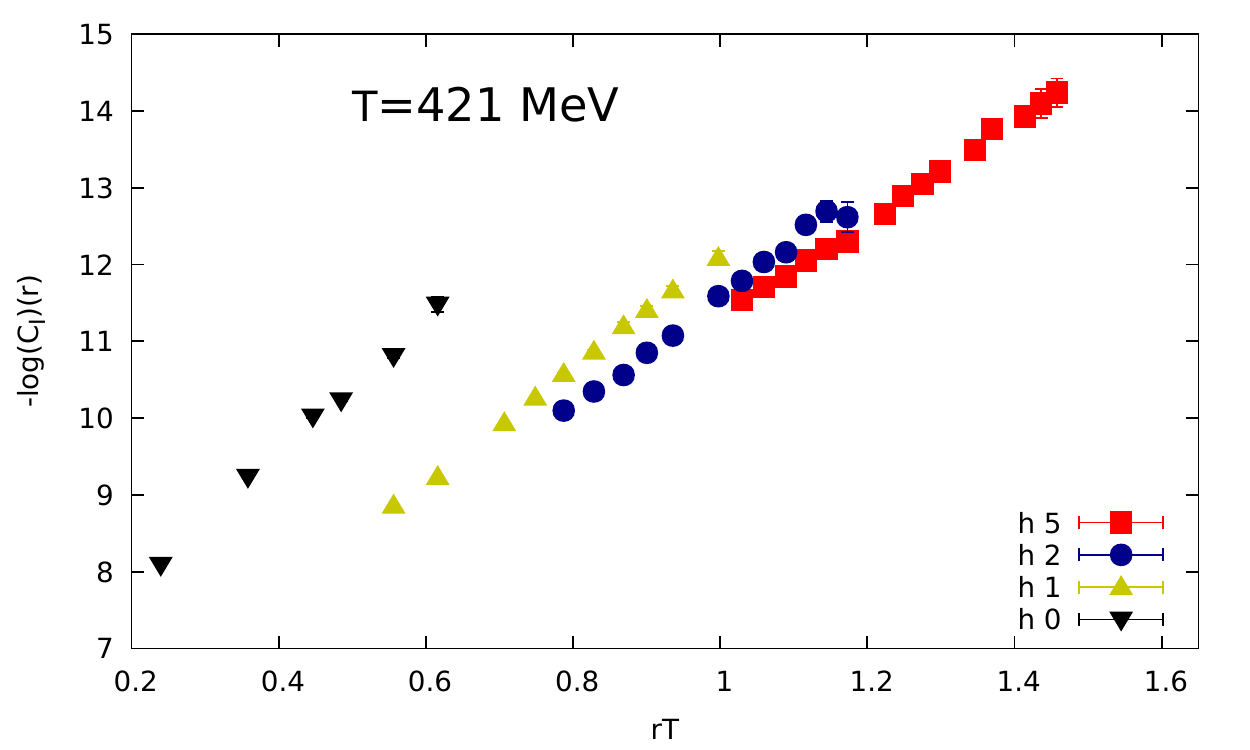}
\hfill
\includegraphics[width=0.49\textwidth]{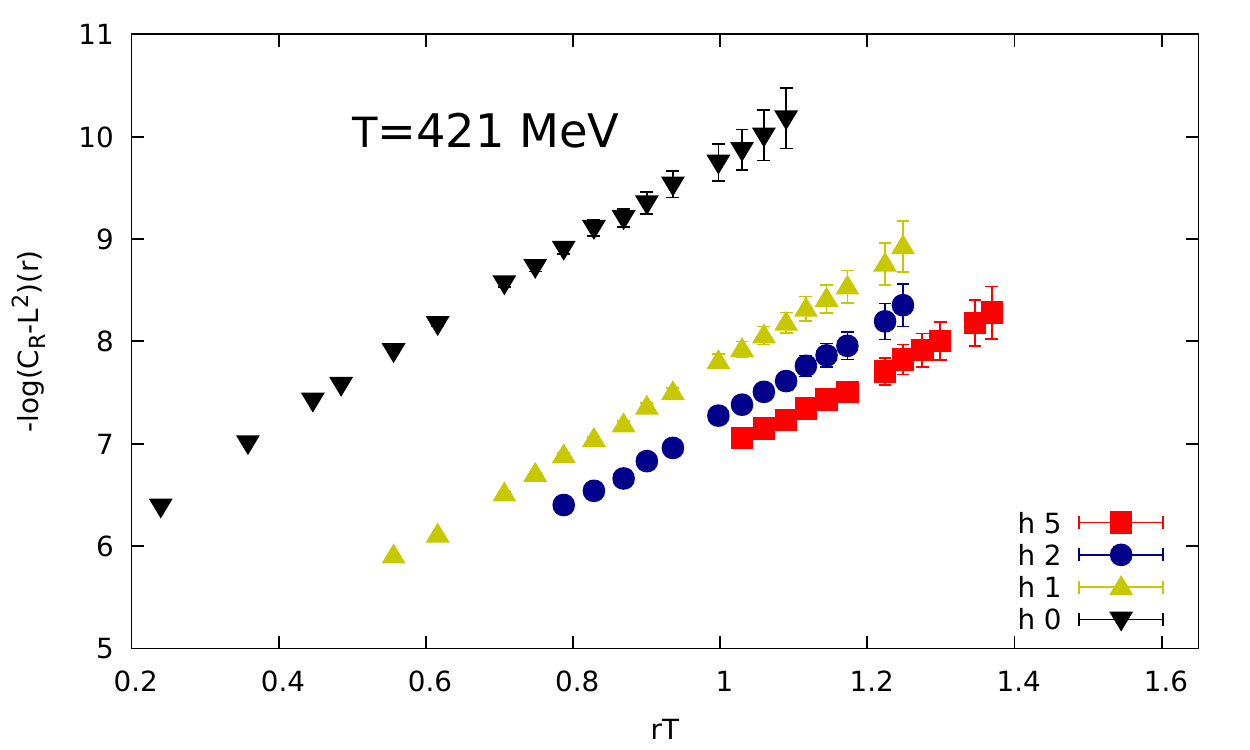}
\caption{Imaginary (left panel) and subtracted real part (right panel) of the 
Polyakov loop correlator as a function of $rT$. We show a high statistics 
result for $\Nt=4$ with aspect ratio 6 for different HYP-smearing levels.}
\label{fig:CRCI-I}
\end{figure}
EQCD predicts that the lowest states contributing to the correlation function 
of the real and imaginary part of the Polyakov loop correlator are bound states 
of 2 and 3 gluons, respectively. The corresponding screening masses, $m_{R}/T$, 
and $m_{I}/T$ in the asymptotic regime should then scale like
\begin{equation}
\frac{m_{R}}{m_{I}} \sim \frac{2}{3} \,.
\end{equation}
This scaling behavior has been seen on the lattice in 
Refs.~\cite{Datta:2002je,Borsanyi:2015yka}. The corresponding correlators of 
the real and imaginary parts of the Polyakov loop, respectively, read
\begin{equation}
C_{R}(r) = \langle \Re P(0) \Re P(r) \rangle \,, \quad\quad C_{I}(r) = \langle 
\Im P(0) \Im P(r) \rangle \,.
\end{equation}
We use the correlation functions with smeared fields and fit $C_{R,I} \sim 
\exp(-m_{R,I}r)/(rT) + \text{const.}$ to probe the mass of the lowest energy 
state that is exchanged and compare these masses to the weak-coupling 
prediction of EQCD. We show an example of the correlators in 
Fig.~\ref{fig:CRCI-I} as a function of $rT$ for $\Nt=4$ with aspect ratio 6 for 
different HYP-smearing levels. Since the Polyakov loop expectation value is 
real in QCD, $C_{R}$ asymptotes to $L^{2}$, which we can subtract, while the 
imaginary part approaches zero asymptotically. In contrast to the color average 
and color singlet screening functions it is not possible to piece together one 
correlator for the real and imaginary parts of the Polyakov loop correlator 
without changing the normalization by hand. For aspect ratio 4 ensembles the 
S/N is usually significantly worse for both correlators. The imaginary part 
correlator has a worse S/N ratio but the real part correlator suffers from 
thermal IR noise due to the subtraction. This can be seen when, e.g., comparing 
the unsmeared results (black symbols).\\
This in general makes it quite difficult to extract screening masses and even 
more complicated to determine their ratio. Our preliminary results on a few 
ensembles with high statistics are reasonably consistent with the EQCD 
prediction and with Ref.~\cite{Borsanyi:2015yka}, i.e., we obtain as our most 
precise result for the ensemble of Fig.~\ref{fig:CRCI-I} $m_{I}/m_{R} \approx 
1.72(4)$, $m_{R}/T \approx 4.3(1)$, and $m_{I}/T \approx 7.4(2)$. Both masses 
tend to decrease at higher temperatures; $m_{I}/T$ appears to decrease more 
strongly.\\
From our results with aspect ratio 6 we conclude that $m_{I}/m_{R}$ can be 
determined with about $10\%$ accuracy at $rT \gtrsim 0.6$, while smaller distances are contaminated by physics of the electric screening regime.

\section{Summary}

We have performed studies of quark-antiquark correlation functions covering a 
wide range of temperatures from below to far above $T_{c}$. Making use of 
HYP-smearing we have been able to obtain a signal for color singlet and color 
average free energies in the screening regime at $T < T_{c}$ and extract 
corresponding screening masses. In comparison with EQCD predictions for the 
screening masses, we obtain reasonable agreement for $T \gtrsim 
400~\text{MeV}$.\\
A second prediction of EQCD is the ratio of the screening masses of the real 
and imaginary part of the Polyakov loop correlator, which we see reasonably 
well confirmed in our measurement for high $T$, although this analysis is still 
ongoing.\\
At present all of our results are preliminary, as they lack continuum 
extrapolation. The full results will be discussed in a future 
publication~\cite{paper2}.

\clearpage

\acknowledgments
This research was supported by the DFG cluster of excellence "Origin and 
Structure of the Universe" 
(\href{www.universe-cluster.de}{www.universe-cluster.de}). The simulations have 
been carried out on the computing facilities of the Computational Center for 
Particle and Astrophysics (C2PAP) and SuperMUC using the publicly available 
MILC code. This work has been supported in part by the U.S Department of Energy 
through grant contract No. DE-SC0012704.\\
The authors would like to thank P. Petreczky for numerous discussions and 
support.\\
S.S. would like to thank N. Brambilla and A. Vairo for support, and TUMGlobal 
for financial support and MSU and BNL for hospitality during a visit.

\bibliographystyle{JHEP}
\bibliography{\Bibliography/library,bib}

\end{document}